%% file: prb.tex
\documentclass[twocolumn,showpacs,prb,superscriptaddress,aps,floatfix]{revtex4-1}

\usepackage{rotating}
\usepackage{amsmath}
\usepackage{amssymb}
\usepackage{color}
\usepackage{graphicx}
\usepackage[active]{srcltx}
\usepackage[sort&compress]{natbib}
\usepackage{amssymb}
\usepackage[dvips]{epsfig}
\usepackage{float}

\include{alias}

\def\ai         {{\it ab-initio}}
\def\ga         {\alpha}
\def\gb         {\beta}
\def\gC         {\Gamma} 
 
\def\gee        {\varepsilon}
\def\gl         {\lambda}

\def\go         {\omega}
\def\goql       {\omega_{\qq \gl}}
\def\gql        {\qq \gl}
\def\gS         {\Sigma}
\def\la         {\langle}
\def\ra         {\rangle}
\def\kk         {{\bf k}}
\def\qq         {{\bf q}}

\def\epc        {EP coupling}
\def\epcp       {EP coupling. }
\def\epi        {EP interaction }
\def\epip       {EP interaction. }
\def\ep         {EP}
\def\dw         {Debye--Waller }

\def\MB         {many body }
\def\se         {self-energy }
\def\sev        {self-energy, }
\def\sep        {self-energy. }
\def\Sf         {SF}
\def\Sfs        {SFs}

\def\nk         {n\kk}

\def\uot        {\frac {1}{2}}

\renewcommand{\[}{\left[}
\renewcommand{\]}{\right]}
\renewcommand{\(}{\left(}
\renewcommand{\)}{\right)}

\newcommand{\grenoble}{Institut Laue Langevin BP 156 38042 Grenoble, France}
\newcommand{\sanseb}{Nano-Bio Spectroscopy Group and
ETSF Scientific Development Centre, Dpto. F\'isica de Materiales,
Universidad del Pa\'is Vasco, Centro de F\'isica de Materiales
CSIC-UPV/EHU-MPC and DIPC, Av. Tolosa 72, E-20018 San Sebasti\'an,
Spain } 
\newcommand{\cnr} {Istituto di Struttura della Materia of the National Research Council, Via Salaria Km 29.3,
I-00016 Monterotondo Stazione, Italy}
\newcommand{\etsf} {European Theoretical Spectroscopy Facilities (ETSF)}

\begin{document}
\title{Ab-initio study of the effects induced by the electron--phonon scattering in carbon based nanostructures}

\author{Elena Cannuccia}
\affiliation{\grenoble} 
\affiliation{\sanseb} 

\author{Andrea Marini}
\affiliation{\cnr} 
\affiliation{\etsf}

\date{\today}
\begin{abstract}
In this paper we investigate from first principles the effect of the electron--phonon interaction in 
two paradigmatic nanostructures: \tpa\, and \tpe.  
We found that the strong electron--phonon interaction leads to the appearance of complex structures
in the frequency dependent electronic self--energy. Those structures rule out any quasi--particle picture,
and make the adiabatic and static approximations commonly used in the well--established Heine Allen Cardona (HAC) approach inadequate.
We propose, instead, a fully \ai\,\,dynamical formulation of the problem within the \MBPT\, framework. 
The present dynamical theory reveals that the structures appearing in the electronic self--energy are 
connected to the existence of packets of correlated  electron/phonon states.
These states appear in the spectral functions even at $T=0\,K$,
revealing the key role played by the zero point motion effect. 
We give a physical interpretation of these states 
by disclosing their internal composition by mapping the \MB
problem to the solution of an eigenvalue problem.
\end{abstract}           


\maketitle

%%%%%%%%%%%%%%%%%%%%%%%%%%%%%%%%%%%%%%%%%%%%%%%%%%%%%%%%%%%%%%%%%%%%%%%%
\section{Introduction}
\label{intro}
%%%%%%%%%%%%%%%%%%%%%%%%%%%%%%%%%%%%%%%%%%%%%%%%%%%%%%%%%%%%%%%%%%%%%%%%%%%%%%%%%%%%%%%%
The electron--phonon\,(EP) coupling is well known to play a key role in several physical phenomena.
For example it affects the renormalization of the electronic bands\cite{allen1983}, the carriers mobility 
in organic devices\cite{GosarChoi1966} or the position and intensity of Raman peaks\cite{attaccalite2010}.
The \epc\, is also the driving force that causes excitons dissociation at the donor/acceptor interface in organic 
photovoltaic\cite{Tamura2008} and the transition to a superconducting phase in solids\cite{supercond}. 

Despite the development of more powerful and efficient computational resources 
the calculation of the effects induced by the EP coupling in realistic materials remains
a challenging task. In addition to the numerical difficulties, it has been assumed, for a long time, that
this interaction can yield only minor corrections (of the order of meV) to the electronic levels.
As a consequence the majority of the \ai\,\,simulations of the electronic and optical properties of
a wide class of materials are generally performed by keeping the atoms frozen in their crystallographic positions. 
It is actually well--known that phonons are atomic vibrations and, as a such, can be easily populated 
by increasing the temperature. This naive observation is {\em de-facto} used to  associate the effect of the EP 
coupling to a temperature effect that vanishes as the temperature goes to zero. However this is not correct
as the atoms posses an intrinsic spatial indetermination due to their quantum nature, that is independent on
the temperature. 
These quantistic oscillations are taken into account by the EP coupling when $T\rar 0$ in the shape of a zero--point--motion effect.

Many years ago\cite{Cardona2006} Heine, Allen and Cardona (HAC) pointed out the EP coupling can induce corrections 
of the electronic levels as large as those induced by the electronic correlation. 
As a consequence the generally accepted statement that the EP coupling always yields minor corrections was doomed to fail.
Nowadays, the advent of more refined numerical techniques, has made possible to ground the HAC approach
in a fully \ai\,\,framework. This has been used to compute the gap renormalization in carbon--nanotubes\cite{capaz2005},
the finite temperature optical properties of semiconductors and insulators \cite{marini2008}, and to confirm a large 
zero--point renormalization ($615$ meV) of the band--gap of bulk diamond \cite{Giustino2010}, previously calculated 
by Zollner using semi--empirical methods\cite{Zollner1992}.
These works are calling into question decades of results, by instilling the doubt that a solely electronic theory may 
be inadequate.  

In this work we show that in nano--structures one of the approximations most commonly used in the electronic theories,
the quasi--particle (QP) approximation\cite{landauqp}, is seriously questioned by the effect of the \epcp 
Indeed in most electronic systems characterized by a moderate internal correlation, the electrons are 
believed to occupy well defined energy levels characterized by a precise energy, width and wave--function. The QP picture
pictorially represents the effect of the correlation on these states as an
electron--hole (in the case of electron--electron coupling) or electron--phonon (in the case of the \epc) pairs cloud
which renormalizes the energy and the width of the electronic level, also reducing its effective electronic charge.
The breakdown of the QP picture caused by the \epc\, has been already
predicted in the case of superconductors by Scalapino et al. \cite{scalapino} and in complex metallic surfaces
by Eiguren et al.\cite{claudia_prl}. More recently we have shown\cite{cannuccia} 
a strong renormalization of the electronic properties of diamond and {\it trans}-polyacetylene caused by the EP coupling in the
zero temperature limit.

In this paper we will extend our previous work\cite{cannuccia}, by providing more methodological and technical details of 
the dynamical theory we have previously used. We will also apply the same method to another polymer, \tpe, finding a severe 
breakdown of the QP picture. The analysis of the \tpe\, results will confirm and strengthen the general conclusions that we 
drew regarding the enormous impact of the electron--phonon coupling in carbon based nano--structures.

In sections \ref{sec:MBPT} and \ref{sec:DW} we will review the derivation of the fully frequency
dependent \se by using the \MBPT. The HAC theory will be, then, found as a static and adiabatic limit of the dynamical 
theory. 
In section \ref{sec:DynamicalSEeffects} and \ref{sec:brakdownQPapprox} we will discuss how the structures appearing in the 
spectral functions of \tpa\,\,and \tpe\,\,rule 
out the basic assumptions of the HAC approach imposing the use of a fully dynamical theory.
In section\,\ref{sec:frohlich} we will show how the problem can be mapped in the solution of a fictitious Hamiltonian that 
makes possible to define the polaronic states as complex electron--phonon packets.
Finally, in the conclusions, we will point out as these results represent an important step forward in the simulation of 
nanostructures, with a wealth of possible implications in the development of more refined theories for the electronic and 
atomic dynamics.

%%%%%%%%%%%%%%%%%%%%%%%%%%%%%%%%%%%%%%%%%%%%%%%%%%%%%%%%%%%%%%%
\section{A dynamical approach to the electron--phonon problem}
\label{sec:MBPT}
%%%%%%%%%%%%%%%%%%%%%%%%%%%%%%%%%%%%%%%%%%%%%%%%%%%%%%%%%%%%%%%
We start from the generic form of the total Hamiltonian of the system that we divide in electronic ($\widehat {H}_{el}$),
atomic ($\widehat{H}_{at}$) and electron--atom part ($\widehat{H}_{el-at}$):
\begin{align}
\widehat{H}=\widehat{H}_{el}+\widehat{H}_{at}+\widehat{H}_{el-at}.
\label{eq:sec_MBPT_1}
\end{align}
The Hamiltonian $\widehat H$ admits both electronic and vibrational states that are coupled by $\widehat H_{el-at}$. 
In this work Density Functional Theory\,(DFT)\cite{R.M.Dreizler1990} is used to calculate
the eigenstates of $\overline{\widehat{H}}$, where we use the 
notation $\overline{O}$ to indicate a quantity or an operator that is evaluated 
with the atoms frozen in their equilibrium crystallographic positions.
Similarly the vibrational states of the Hamiltonian $\widehat H$ are described, fully \ai, 
by using the well--known extension of DFT, the Density Functional Perturbation Theory\,(DFPT)\cite{baroni2001,Gonze1995}. 
In DFPT the electronic correlations are embodied in a self--consistent mean potential $\widehat V_{scf}$ representing the
{\it total} electronic potential which depends on the atomic positions $\RR_{Is}\equiv \RR_I+\tau_s$:
\begin{align}
\widehat{H}_{el-at}=\int_{crystal}\,d\rr\, \hat{\gr}\(\rr\) \Vscf\[\{\RR\}\]\(\rr\).
\label{eq:sec_MBPT_2}
\end{align}
In the definition of $\RR_{Is}$, $I$ and $s$ label the lattice cell (at position $\RR_I$) and the atoms in the cell
(at position $\tau_s$), respectively. In Eq.(\ref{eq:sec_MBPT_2}) $\gr$ is the electron density operator.

The aspect we are interested in this paper is how to properly include the modifications of the
electronic levels induced by the atomic vibrations. In particular, by assuming the harmonic  approximation
for the phonons, we will develop a dynamical theory of the electronic dynamics.
To this end we follow a purely diagrammatic approach\cite{mahan} 
to present a short but accurate review of the derivation of the Fan\cite{fan1951} self--energy
and of the much less known \dw (DW) correction\cite{marini_2012}.

If we now consider a configuration of lattice displacements $\hat{\bf u}_{Is}$, $\widehat H$ can be expressed as a Taylor expansion
\begin{multline}
\widehat{H}-\overline{\widehat{H}} 
= \sum_{I s \ga} \overline{\frac{\partial \Vscf\[\{\RR\}\]\(\rr\)}{\partial{R_{Is\ga}}}} 
\hat{u}_{I s \ga}+\\+\frac{1}{2}
\sum_{I s \ga, J s' \gb} \overline{\frac{\partial^2 \Vscf\[\{\RR\}\]\(\rr\)}{\partial{R_{Is\ga}}\partial{R_{Js'\gb}} }} \hat{u}_{I s \ga}\hat{u}_{J s'
\gb},
\label{eq:sec_diagrams_1}
\end{multline}
where $\ga$ and $\gb$ are the Cartesian coordinates.

The link with the perturbative expansion is readily done by transforming Eq.\,(\ref{eq:sec_diagrams_1}) from the space of 
the lattice displacements to the space of the canonical lattice vibrations by means of the identity\cite{mattuck}: 
\begin{multline}
\hat{u}_{I s \ga}=\sum_{\qq \gl} \(2 N_q M_s \go_{\qq \gl}\)^{-1/2} \xi_{\ga}\(\qq \gl|s\) e^{i \qq\cdot\(\RR_I+\tau_s\)}\times\\
\times\(\hat{b}^{\dagger}_{-\qq \gl}+\hat{b}_{\qq \gl}\),
\label{eq:sec_diagrams_2}
\end{multline}
where $N_q$ is the number of cells (or, equivalently the number of q--points) used in the simulation and $M_s$ is the atomic mass
of the $s$ atom in the unit cell. $\xi_{\ga}\(\qq \gl|s\)$ is the phonon polarization vector
and $\bdql$ and $\bql$ are the bosonic creation and annihilation operators.

By inserting Eq.\,(\ref{eq:sec_diagrams_2}) into Eq.\,(\ref{eq:sec_diagrams_1}) we get
\begin{multline}
\widehat{H}-\overline{\widehat{H}} 
= \frac{1}{\sqrt{N_q}}\sum_{\kk n n^{'} \qq \gl} g^{\gql}_{n n' \kk} 
\hat{c}^{\dagger}_{n\kk} \hat{c}_{n'\kk-\qq} \(\hat{b}^{\dagger}_{-\qq \gl} +\hat{b}_{\qq \gl} \)+\\
+\frac{1}{N_q}\sum_{n n^{'}\kk}\sum_{\qq \gl, \qq' \gl'} \Lambda^{\gql,\qq'\gl'}_{nn' \kk} c^{\dagger}_{n\kk} c_{n'\kk-\qq-\qq'} \times\\
\times\(\hat{b}^{\dagger}_{-\qq \gl} +\hat{b}_{\qq \gl} \)
\(\hat{b}^{\dagger}_{-\qq' \gl'} +\hat{b}_{\qq' \gl'} \).
\label{eq:sec_diagrams_3}
\end{multline}
In Eq.\,(\ref{eq:sec_diagrams_3}) we have introduced the first--order ($g^{\gql}_{n' n \kk}$) and the second--order
($\Lambda^{\gql,\qq'\gl'}_{n' n \kk}$) electron--phonon matrix elements which will be shortly defined. 
To this purpose we rewrite $\widehat V_{scf}$ making explicit its dependence on the atomic positions: 
\begin{align}
\Vscf\[\{\RR\}\]\(\rr\)=\sum_{Is} \Vscf\(\rr-\RR_{Is}\).
\label{eq:sec_diagrams_3a}
\end{align}
From Eq.\,(\ref{eq:sec_diagrams_3a}) it follows that the
second order derivatives in the atomic positions are diagonal, 
$\frac{\partial^2}{\partial \RR_{Is} \partial \RR_{Js'}}\Vscf\[\{\RR\}\]\(\rr\)\propto 
\gd_{IJ}\gd_{ss'} \frac{\partial^2}{\partial \RR^2_{Is}}\Vscf\[\{\RR\}\]\(\rr\) $. 

By using Eq.\,(\ref{eq:sec_diagrams_3a}) the summation on $\RR_{I}$ appearing in Eq.\,(\ref{eq:sec_diagrams_1}) leads to 
the momentum conservation both in the first and second order terms. At the first order this leads to the definition of 
the electron--phonon matrix elements
\begin{multline}
g^{\gql}_{n n' \kk}=\sum_{s \ga} \(2 M_s \go_{\gql}\)^{-1/2} e^{i\qq\cdot\tau_s} \times \\
\times \la n\kk |\frac{\partial \Vscf^{\(s\)}\(\rr\)}{\partial{R_{s\ga}}} | n' \kk-\qq \ra \xi_{\ga}\(\qq \gl|s\),
\label{eq:sec_diagrams_4}
\end{multline}
We have also used the short 
form $R_{s \ga}=\left. R_{Is\ga}\right|_{I=0}$.
A similar derivation can be followed to derive the 2$^{nd}$ order term
\begin{multline}
    \Lambda^{\gql,\qq'\gl'}_{n n' \kk}= \frac{1}{2}\sum_{s}\sum_{\ga,\gb} \frac{ \xi^{*}_{\ga}\(\qq \gl|s\) 
    \xi_{\gb}\(\qq' \gl'|s\)}
{2M_s\(\go_{\gql} \go_{\qq' \gl'} \)^{1/2}} \times \\
\times \la n\kk |\frac{\partial^2 \Vscf^{\(s\)}\(\rr\)}{\partial{R_{s\ga}}\partial{R_{s\gb}}} | n' \kk-\qq-\qq' \ra.
\label{eq:sec_diagrams_5}
\end{multline}
This second--order term is, in general, 
neglected as it is assumed to be small compared to the first--order term. Although this is correct at the level of the 
Hamiltonian it is not true anymore even at the lowest order of perturbation theory.

\begin{figure}[h]
\begin{center}
\parbox[c]{4cm}{
\begin{center}
\epsfig{figure=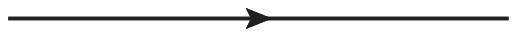,width=3cm}\\
${\cal G}_{n \kk}^{(0)}(\gon)$
\end{center}
}
\parbox[c]{4cm}{
\begin{center}
\epsfig{figure=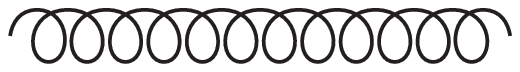,width=3cm}\\
${\cal D}_{\qq \gl}^{(0)}(\goj)$
\end{center}
}\\
\parbox[c]{4cm}{
\begin{center}
\epsfig{figure=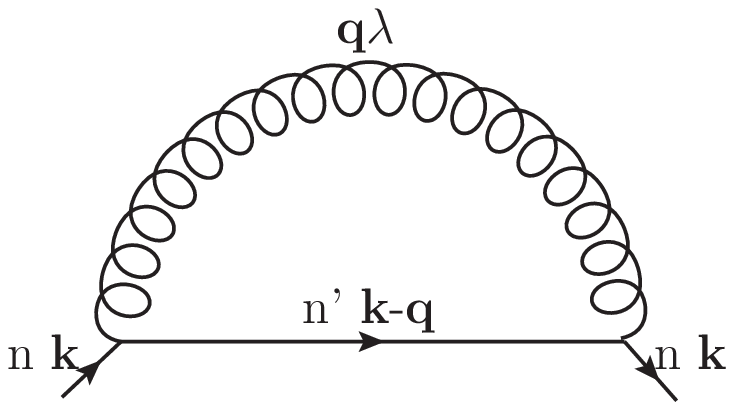,width=4cm}\\
\text{(a) 2$^{nd}$ order $\Sigma^{Fan}$}
\end{center}
}
\parbox[c]{4cm}{
\begin{center}
\epsfig{figure=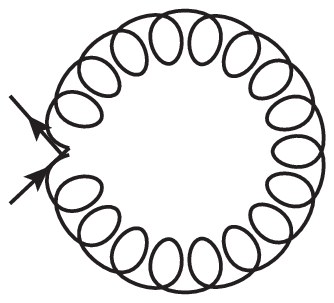,width=3cm}\\
\text{(b) 2$^{nd}$ order $\Sigma^{DW}$}
\end{center}
}\\
\parbox[c]{4cm}{
\begin{center}
\epsfig{figure=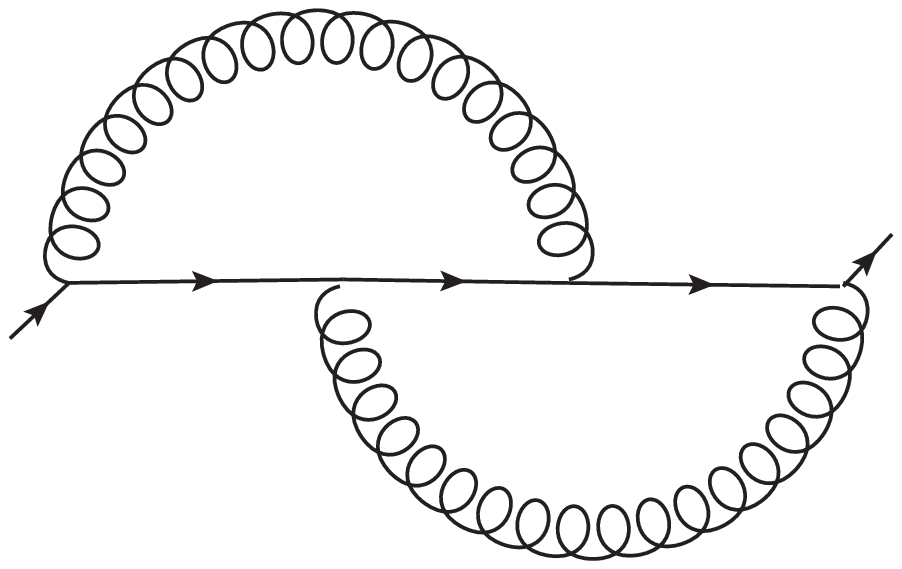,width=4cm}\\
\text{(c) 4$^{th}$ order $\Sigma^{Fan}$}
\end{center}
}
\parbox[c]{4cm}{
\begin{center}
\epsfig{figure=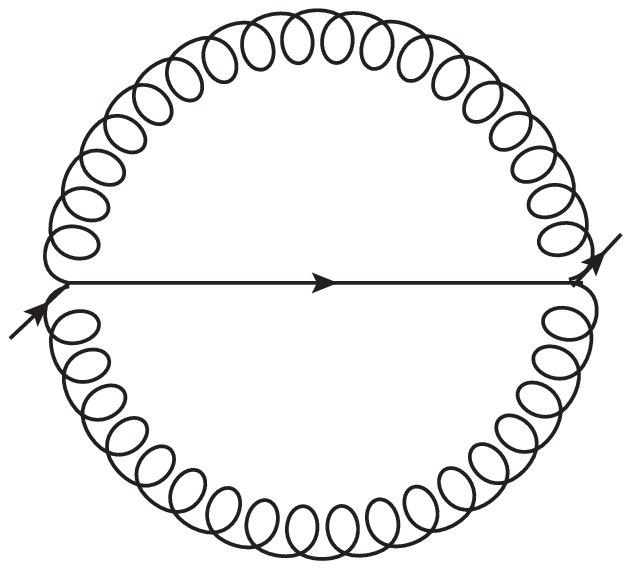,width=3cm}\\
\text{(d) 4$^{th}$ order $\Sigma^{DW}$}
\end{center}
}
\end{center}
\caption{\footnotesize{
The \se diagrams corresponding to the first and second order terms in the Taylor expansion of $\widehat{H}-\overline{\widehat{H}}$ (see Eq.\,(\ref{eq:sec_diagrams_1}))
treated at different orders of the perturbative expansion. 
For example the well--known
Fan \se is formally obtained as a 2$^{nd}$ order expansion of the first order term in Eq.\,(\ref{eq:sec_diagrams_1}).
However the second term of Eq.\,(\ref{eq:sec_diagrams_1}), treated at first order gives the 2$^{nd}$ order $\Sigma^{DW}$, that is of the
same order of the Fan term and, consequently, cannot be neglected. The diagram (c) is obtained as a $4^{th}$ order expansion of the first order 
in Eq.\,(\ref{eq:sec_diagrams_1}) while (d) comes from the second term of Eq.\,(\ref{eq:sec_diagrams_1}) treated at the $2^{th}$ order.
}}
\label{fig:sec_diagrams_1}
\end{figure}

Indeed the different terms in the Taylor expansion of the Hamiltonian defined in Eq.\,(\ref{eq:sec_diagrams_3}) 
induce a wealth of diagrams of increasing complexity and order.
If we restrict to the lowest non vanishing order we have two diagrams: the Fan\cite{fan1950} and
the DW. These are presented by diagrams $(a)$  and $(b)$ in Fig.\,(\ref{fig:sec_diagrams_1}). In the same
Figure two fourth order (in the displacements) diagrams, ($c$) and ($d$),
are also showed. They are of the same order and, as the Fan and DW diagrams, they result from the perturbative treatment of 
the first order and second order terms in Eq.\,(\ref{eq:sec_diagrams_3}).

The actual calculation of the Fan and DW diagrams is straightforward. The Fan's diagram is similar to the one generated by the electronic correlation
in the so-called GW approximation\cite{strinati}, where the screened electronic interaction is replaced by a phonon 
propagator of wave vector $\qq$ and branch $\gl$\cite{mahan}.
Applying the finite temperature diagrammatic rules it is possible to 
define the Fan \se operator $\Sigma^{Fan}_{n\kk}\(i\goi,T\)$, recovering the expression originally evaluated by Fan\cite{fan1950}:
\begin{multline}
\Sigma^{Fan}_{n\kk}\(i\goi,T\)=-\uob \uon \sum_{\qq \gl} \sum_{n'} \gsq \times\\
\times\sum^{+\infty}_{j=-\infty} 
D^{(0)}_{\gql}\(i\goj\) G_{n'\kk-\qq}^{(0)}\(i\goi-i\goj\),
\label{eq:sec_diagrams_7}
\end{multline}
where $\beta=\frac{1}{\mathit{K}T}$ ($\mathit k$ is the Boltzmann constant) and $T$ is the temperature of the phonon bath.
By using the standard definitions of the electronic and the phononic Green's functions:
$G_{n'\kk}^{(0)}\(i\goi\)=\(\igoi-\gee_{n\kk}+\mu\)^{-1}$ (where $\mu$ is the chemical potential), 
$D_{\gql}^{(0)}\(i\goj\)=\(\(\igoj-\goql\)^{-1}-\(\igoj+\goql\)^{-1}\)$,
and summing over the Matsubara frequencies, we get the final expression for the Fan self--energy
\begin{multline}
    \Sigma^{Fan}_{n\kk}\(i\go,T\) = \sum_{n'\gql} \frac {\gsq}{N_q} \times \\
\times\[ \frac{N_{\qq\gl}\(T\)+1-f_{n'\kk-\qq}}{i\go-\gee_{n' \kk-\qq} -\goql -i0^{+}} \right. + \\
+ \left. \frac{N_{\qq\gl}\(T\)+f_{n' \kk-\qq}}{i\go-\gee_{n' \kk-\qq}+\goql -i0^{+}}\],
\label{eq:sec_diagrams_10}
\end{multline}
where $N_{\qq\gl}\(T\)$ is the Bose function distribution of the phonon mode $\(\qq,\gl\)$ at temperature $T$.

A similar expression can be derived for the frequency independent DW self--energy $\gS_{\nk}^{DW}$. This term comes from
the equal time contractions of the $\(\hat{b}^{\dagger}_{-\qq \gl} +\hat{b}_{\qq \gl} \) \(\hat{b}^{\dagger}_{-\qq' \gl'} +\hat{b}_{\qq' \gl'} \)$
operators:
\begin{multline}
\la \(\hat{b}^{\dagger}_{-\qq \gl} +\hat{b}_{\qq \gl} \) \(\hat{b}^{\dagger}_{-\qq' \gl'} +\hat{b}_{\qq' \gl'} \) \ra =\\
\gd_{-\qq,\qq'}\gd_{\gl,\gl'} \[N_{\qq'\gl}\(T\)+N_{\qq\gl}\(T\)+1\].
\label{eq:sec_diagrams_11}
\end{multline}
The corresponding diagram ($b$) in Fig.\,\ref{fig:sec_diagrams_1} can be easily found to be
\begin{align}
\gS^{DW}_{\nk}\(T\)=\frac{1}{N_q}\sum_{\gql} \Lambda^{\qq\gl,-\qq\gl}_{n n \kk}  \(2 N_{\qq\gl}\(T\) +1\).
\label{eq:sec_diagrams_12}
\end{align}
Both the Fan and DW \se have been already derived previously in the framework of the
Heine--Allen--Cardona\,(HAC) theory~\cite{allen1976,allen1983,Cardona2006}. The HAC approach is based on the 
static Rayleigh-Schr\"{o}dinger perturbation theory. More precisely the $\hat{u}_{Is\ga}$ are used
as scalar variables on which a static perturbation theory is applied.
As we will mention in Sec.\,\ref{sec:DW}, the second--order derivatives appearing in the definition of the DW term can be
rewritten in terms of the one--order derivatives by imposing the translational invariance of the correction to the electronic levels.

On the other hand the Fr\"{o}hlich and the Holstein Hamiltonians usually neglect the DW term (diagram ($b$) in Fig.\,\ref{fig:sec_diagrams_1}), 
even if it is of the same order of the Fan term.  

The \MB formulation represents the dynamical extension of the HAC approach, that is recovered 
from Eq.\,(\ref{eq:sec_diagrams_10}) by using $\go\approx \gee_{n\kk}$ (the on--the--mass--shell\,(OMS) limit) and 
$\left|\gee_{n\kk}-\gee_{n' \kk-\qq}\right|\gg \go_{\qq\gl}$ (the adiabatic limit) and by considering only the real part of the self--energy. 
It turns out, therefore, that in the HAC approach the temperature dependent change in the single--particle energies is given by
\begin{align}
\Delta \gee^{HAC}_{n\kk}\(T\)=\gS^{DW}_{\nk}(T)+
\sum_{n'\gql} \frac {\gsq}{N_q}  \frac{2 N_{\qq\gl}\(T\)+1}{\gee_{n\kk}-\gee_{n' \kk-\qq} }.
\label{eq:sec_diagrams_13}
\end{align}
The soundness of the HAC approach is then, from a MBPT perspective, connected to the validity of the 
on--the--mass--shell and of the adiabatic approximations.
We will prove in the section\,\ref{sec:DynamicalSEeffects} that these 
approximations are not always 
well motivated. Dynamical and non--adiabatic corrections can be
huge, such to invalidate the applicability of the HAC approach.

%%%%%%%%%%%%%%%%%%%%%%%%%%%%%%%%%%%%%%%%%%%%%%%%%%%%%%%%%%%%%%%%%%%%%%%%%%%%%%%%%%%%
\section{Second order derivatives of $\widehat V_{scf}$ and the actual calculation of the Debye Waller \se}
\label{sec:DW}
%%%%%%%%%%%%%%%%%%%%%%%%%%%%%%%%%%%%%%%%%%%%%%%%%%%%%%%%%%%%%%%%%%%%%%%%%%%%%%%%%%%%

The general expression, Eq.\,(\ref{eq:sec_diagrams_3}) and Eq.\,(\ref{eq:sec_diagrams_5}) for the perturbed Hamiltonian requires the knowledge of
the second--order gradients of the self--consistent potential
\begin{align}
\Delta^{s\ga\gb}_{n'\pp,n\kk}=
\la n'\pp |\frac{\partial^2 \Vscf^{\(s\)}\(\rr\)}{\partial{R_{s\ga}}\partial{R_{s\gb}}} | n \kk \ra,
\label{eq:sec_DW_1}
\end{align}
where $\pp$ here replaces the $\kk+\qq+\qq'$ vector appearing in Eq.\,(\ref{eq:sec_diagrams_3}) and Eq.\,(\ref{eq:sec_diagrams_5}).
These terms are extremely cumbersome to calculate and the task becomes easily prohibitive when higher
orders are included. Their evaluation is, however, crucial because the $\kk=\pp$ case is needed to calculate 
the lowest--order DW \sev while the finite momenta matrix element defines higher order 
diagrams (like diagram $(d)$ in Fig.\,(\ref{fig:sec_diagrams_1})).

In the case of the simpler $\gS^{DW}_{\nk}$ (diagram $(b)$ in Fig.\,(\ref{fig:sec_diagrams_1})) we know, from Eq.\,(\ref{eq:sec_diagrams_5}) and Eq.\,(\ref{eq:sec_DW_1}) that
\begin{multline}
\Lambda^{\qq\gl,-\qq\gl}_{n n \kk}= \frac{1}{2}\sum_{s}\sum_{\ga,\gb} \(2 M_s \go_{\gql} \)^{-1} \times \\
 \times \frac{ \xi^{*}_{\ga}\(\qq \gl|s\) \xi_{\gb}\(-\qq \gl|s\)}
{2M_s\go_{\gql}}
\Delta^{s\ga\gb}_{n\kk,n\kk}.
\end{multline}
In order to evaluate the $\Delta^{s\ga\gb}_{n'\pp,n\kk}$ factors
the HAC theory uses the fact that if all atoms were shifted by the same amount all physical
quantities should not change.
In other terms, being the $\Delta \gee^{HAC}_{n\kk}\(T\)$ an explicit functional of the atomic positions (Eq.\,(\ref{eq:sec_diagrams_3a}))
that are treated classically, it is possible to impose the following translational invariance condition
\begin{align}
\Delta \gee^{HAC}_{n\kk}\[\{u_{Is\ga}\}\]\(T\)=\Delta \gee^{HAC}_{n\kk}\[\{u_{Is\ga}+d_{\ga}\}\]\(T\).
\label{eq:sec_DW_2}
\end{align}
From this condition it follows that~\cite{allen1976,allen1983,Cardona2006} in order to calculate
$\Lambda^{\qq\gl,-\qq\gl}_{n n \kk}$ that defines the DW \se (see Eq.\,(\ref{eq:sec_diagrams_12}))
only the matrix element $\Delta^{s\ga\gb}_{n\kk,n\kk}$ is needed. This can be rewritten as
\begin{multline}
\Delta^{s\ga\gb}_{n\kk,n\kk}=
-\sum_{n'\neq n}\frac{1}{\gee_{n\kk}-\gee_{n'\kk}} \times\\  
\times \[ \(\sum_{s'} \la n\kk |\frac{\partial \Vscf^{\(s'\)}\(\rr\)}{\partial{R_{s'\ga}}} | n' \kk \ra\)
 \la n'\kk |\frac{\partial \Vscf^{\(s\)}\(\rr\)}{\partial{R_{s\gb}}} | n \kk \ra
\right. + \\\left.
+ \la n\kk |\frac{\partial \Vscf^{\(s\)}\(\rr\)}{\partial{R_{s\ga}}} | n' \kk \ra
\(\sum_{s'} \la n'\kk |\frac{\partial \Vscf^{\(s'\)}\(\rr\)}{\partial{R_{s'\gb}}} | n \kk \ra\)
\].
\label{eq:sec_DW_3}
\end{multline}
The condition given by Eq.\,(\ref{eq:sec_DW_2}) is, however, 
intrinsically ill--defined in the diagrammatic approach: the correction to the energy levels
is a quantity obtained in fact, from the \se operator that, in turns, can be defined {\em only} when the displacement operators
are quantized and a second quantized form of the Hamiltonian change (Eq.\,(\ref{eq:sec_diagrams_3})) is introduced.

This inconsistency can be, indeed, cured in a fully MBPT framework~\cite{marini_2012} but it requires to introduce the constant displacement
vector $d_\ga$ as an operator. This leads to the definition of new kinds of diagrams that will depend on powers of $d_\ga$. By imposing that diagrams of 
the same order cancel each other it is possible to obtain a general expression for the matrix element
$\Delta^{s\ga\gb}_{n'\pp,n\kk}$.

As discussed by X. Gonze\cite{gonze}, the local dependence on the atomic positions in Eq.\,(\ref{eq:sec_diagrams_3a}) assumes that the
electronic screening of the ionic potential, that defines $\Vscf$, depends only smoothly on $\RR_{Is}$. By taking fully into account this intrinsic
dependence on the atomic positions a correction to the Debye--Waller term, named non--diagonal Debye--Waller correction, can be defined.
This correction has been reported to be important for isolated molecules and atoms\cite{gonze}. Its effect in solids and, more generally,
in extended systems is expected to be weakened by the efficient screening properties.

%%%%%%%%%%%%%%%%%%%%%%%%%%%%%%%%%%%%%%%%%%%%%%%%%%%%%%%%%%%%%%%%%%%%%%%%%%%%%%%%%%%%
\section{Dynamical Self-Energy Effects beyond the Quasi Particle Approximation}
\label{sec:DynamicalSEeffects}
%%%%%%%%%%%%%%%%%%%%%%%%%%%%%%%%%%%%%%%%%%%%%%%%%%%%%%%%%%%%%%%%%%%%%%%%%%%%%%%%%%%%
In section \ref{sec:MBPT} we showed that the HAC theory represents the static and adiabatic limit of the dynamical 
electron--phonon \sep
The more suitable are the conditions of validity of the on--the--mass--shell and of the adiabatic approximations, the sounder is the 
applicability of the HAC approach from a MBPT perspective.
We want to prove in this section that these approximations are not always well motivated and dynamical and 
non--adiabatic corrections to the HAC approach cannot be neglected {\it a priori}.

The HAC approach grounds on the concept of a well defined QP state: the charge carriers are assumed to be 
concentrated on electronic levels, being characterized by a well defined energy and wave-function. The QP concept can be firmly
introduced in a many-body \GF theory\,\cite{mahan} where the definition embodies, at the same time, its limitations, as it will be 
clear in the following.

The fully interacting propagator can be written, for real energies $\go$ 
in terms of the \sev (Eq.\,(\ref{eq:sec_diagrams_7})) as
\begin{equation}
G_{\nk}\(\go,T\)=\frac {1}{\go-\gee_{\nk}-\gS^{Fan}_{\nk}\(\go,T\)-\gS^{DW}_{\nk}\(T\)}.
\label{eq:Dyson_withSE}
\end{equation}
The rotation from the imaginary to the real axis has been easily performed by a Wick rotation\cite{mattuck} 
as the energy dependence of the Fan self--energy is explicit.
The single particle excitations are then the complex poles of Eq.\,(\ref{eq:Dyson_withSE}) 
\begin{multline}
\go-\gee_{\nk}-\Sigma^{DW}_{\nk}\(T\)-\Re \[\gS^{Fan}_{\nk}\(E_{\nk}\(T\),T\)\]+\\
-i\Im \[\gS^{Fan}_{\nk}\(E_{\nk}\(T\),T\)\]=0.
\label{eq:FindingPoles}
\end{multline}
As it is clear from Eq.\,(\ref{eq:FindingPoles}) a genuine QP state should have a zero line-width, that is a zero 
imaginary part of the \sep In practice this is never completely true.
Nevertheless, when the frequency dependence of the \se is smooth Eq.\,(\ref{eq:Dyson_withSE}) can be rewritten by using two simple
and intuitive approximations: the OMS and the QP approximation. In the specific case of a constant and real \se
(as in the HAC case) one can introduce the OMS approximation where the solution of Eq.\,(\ref{eq:FindingPoles}) is given by
\begin{align} 
E_{\nk}\(T\) = \gee_{\nk} + \Sigma^{DW}_{\nk}\(T\) + \gS^{Fan}_{\nk}\(\gee_{\nk},T\).
\label{eq:energy_OMS}
\end{align} 

We notice, from Eqs.\,(\ref{eq:FindingPoles}) and (\ref{eq:energy_OMS}), that $E_{\nk}\(T\)$ is complex.
Even in the case where the \se is not constant, if the bare energy $\gee_{\nk}$ is far from a pole of \sev
then $\gS^{Fan}_{\nk}\(\omega,T\)$ can be Taylor expanded, up to the first order, around $\gee_{\nk}$:
\begin{multline}
E_{\nk}\(T\) = \gee_{\nk}+\Sigma^{DW}_{\nk}\(T\)+\gS^{Fan}_{\nk}\(\gee_{\nk},T\)+\\
+\left. \frac{\partial \gS^{Fan}_{\nk}\(\go,T\)}{\partial \go}\right|_{\go=\gee_{\nk}}\(E_{\nk}\(T\)-\gee_{\nk}\).
\label{eq:QP_newton}
\end{multline}
Eq.\,(\ref{eq:QP_newton}) corresponds to the QP approximation.
The bare energy is then renormalized because of the virtual 
scatterings which are described by the real part of the \sep This renormalization is easily
described by the solution of Eq.\,(\ref{eq:QP_newton}):
\begin{multline}
E_{\nk}\(T\) = \gee_{\nk} + \\+ Z_{\nk}\(T\) \[\gS^{Fan}_{\nk}\(\gee_{\nk},T\)+\Sigma^{DW}_{\nk}\(T\)\],
\label{eq:QP_energy}
\end{multline}
with $Z_{\nk}\(T\) = \(1-\left. \frac{\partial \gS^{Fan}_{\nk}\(\go,T\)}{\partial \go}\right|_{\go=\gee_{\nk}}\)^{-1}$ the 
renormalization factor.
From Eq.\,(\ref{eq:QP_energy}) it is evident that $E_{\nk}$ is complex and its imaginary part $\Gamma_{\nk}\(T\)=\Im\[E_{\nk}\(T\)\]$, the QP line-width, is proportional
to the  $\Im \[ Z_{\nk}\(T\)\gS^{Fan}_{\nk} \(\gee_{\nk},T\)\]$. A small $\Gamma_{\nk}\(T\)$ 
indicates a stable QP, that slowly decays because of the real scatterings with the other particles
and with the phonon modes.
By assuming the QP approximation to be valid Eq.\,(\ref{eq:Dyson_withSE}) can be re-written as $G_{\nk}\(\go,T\)=Z_{\nk}\(T\)\(\go-E_{\nk}\(T\)\)^{-1}$.
\begin{figure} [h]
\begin{center}
\parbox[c]{7cm}{
\begin{center}
\epsfig{figure=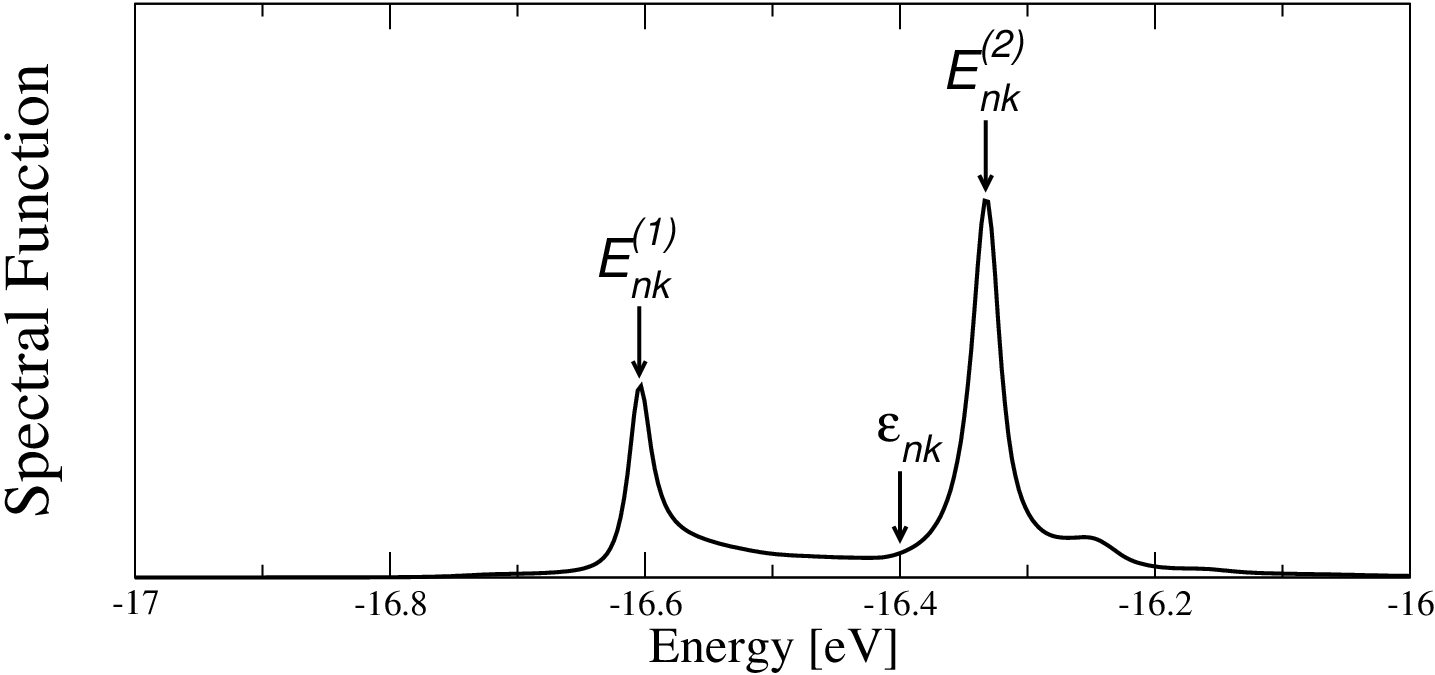,width=7.0cm,clip=,bbllx=28,bblly=200,bburx=715,bbury=530} 
\end{center}
}\\[-.5cm]
\parbox[c]{7cm}{
\begin{center}
\epsfig{figure=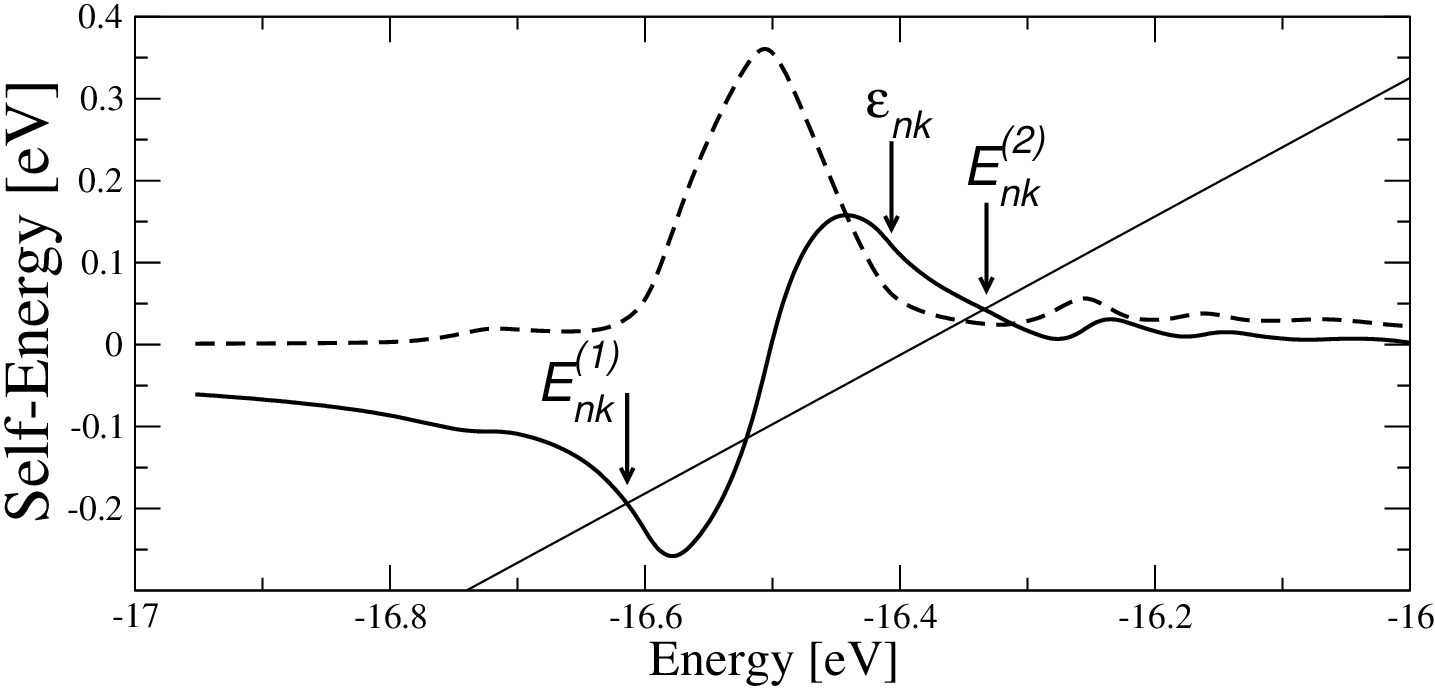,width=7.0cm,clip=,bbllx=28,bblly=200,bburx=715,bbury=530} 
\end{center}
}
\end{center}
\vspace{-.2cm}
\caption{\footnotesize{\Tpa. Spectral function (upper frame) and self--energy (lower frame) corresponding to the state $\mid n=1,\kk=\gC \ra$.
In the self--energy frame both the real (solid line) and imaginary (dashed line) parts are showed. The three arrows represent the
bare electronic energy $\gee_{\nk}$ and the two solutions ($E^{\(1\)}_{\nk}$,$E^{\(2\)}_{\nk}$) of Eq.\,(\ref{eq:FindingPoles}). 
The thin solid straight line represents instead the function $\go-\gee_{\nk}-\Sigma_{\nk}^{DW}$.
As the imaginary part of the self--energy shows a clear, intense and wide peak at around $-16.5$\,eV (i.e. very close to $\gee_{\nk}$) 
the real--part is dominated by a rapid oscillation that cannot be captured at all by the linearization of the energy dependence and 
causes the appearance of two solutions of Eq.\,(\ref{eq:FindingPoles}).}}
\label{fig:SF_b1k1}
\end{figure}

Angle Resolved Photoemission Spectroscopy (ARPES) provides a definitive tool to verify if the QP approximation is accurate. 
If it existed a true QP should appear as a peak in the
photoemission spectra with a Lorentzian lineshape. Indeed one finds that the spectral function\,(SF)
$A_{\nk}\(\go,T\)\equiv\pi^{-1}\mid\Im\[G_{\nk}\(\go,T\)\]\mid$
is given, in the QP approximation and in the simple case of a purely real $Z_{\nk}$, by
\begin{align}
A^{\(qp\)}_{n\kk}\(\go,T\)=\frac{Z_{\nk}\(T\)|\Gamma_{\nk}\(T\)|}{\pi\[\(\go-\Re\[E_{\nk}\(T\)\]\)^2+\Gamma^2_{\nk}\(T\)\]}.
\label{eq:SQ_qp}
\end{align}
In this case the QP energy and width give the peak position and the spectral peak width.
It is worth noticing that a complex value of $Z_{\nk}\(T\)$ would cause the spectral function to have an asymmetric lineshape.

The SF gives a physical interpretation and a clear validation of the QP approximation. 
Indeed $A^{\(qp\)}_{n\kk}\(\go,T\)$ is a probability function to find an electron in the state $\nk$ with energy $\go$
and the total electronic charge associated to the QP state
is $Z_{\nk}\(T\)$, that corresponds to the integral of $A^{\(qp\)}_{\nk}\(\go,T\)$. The renormalization factor represents, therefore, the QP charge.
When $Z_{\nk}\(T\)=1$ and $\Im\[ \gS_{\nk}(\gee_{\nk},T)\]\rar 0$ the SF reduces to a delta function, 
the SF of a particle with energy $\Re\[E_{\nk}\(T\)\]$. 

It is clear that a direct comparison of $A^{\(qp\)}_{n\kk}\(\go,T\)$ with the true SF corresponding to a given self--energy
or with the ARPES lineshape provides the ultimate validation of the QP picture. A paradigmatic example that well explains the basic
mechanism for the breakdown of the QP picture is given in Fig.\,\ref{fig:SF_b1k1} where the zero temperature SF for the 
 $\mid n=1,\kk=\gC \ra$ state of \tpa\,\,is showed in the upper frame. It is clear that the SF of this state is far
from being well represented by a Lorentzian lineshape. Indeed it is evident the appearance of two 
peaks at energies $E^{\(1\)}_{\nk}$ and $E^{\(2\)}_{\nk}$. 
These two peaks are the signature of a breakdown of the QP picture, because the existence condition of only one pole collecting most of the weight is not satisfied. 
We will come back on the physical interpretation of such structures in the next section.
Nevertheless the origin of these two peaks is evident if we 
analyze the energy dependence of the imaginary and real parts of the corresponding self--energy, showed in the lower frame of the same
figure. In this specific case the solution of Eq.\,(\ref{eq:FindingPoles}) admits two roots as a consequence of the rapid oscillation
of $\Re\[\Sigma^{Fan}_{\nk}\(\go,T=0\)\]$ around $-16.5$\,eV (the bare electronic energy of this state). 
This oscillation is, in turn, induced by an intense peak appearing in the
$\Im\[\Sigma^{Fan}_{\nk}\(\go,T=0\)\]$. 
We deduce that in this case a naive application of the QP approximation in form of a linearization of 
$\Sigma^{Fan}_{\nk}\(\go, T\)$ (Eq.\,(\ref{eq:QP_newton}))
may produce a non physical energy dependence of the self--energy. As a consequence this leads to meaningless (negative or enormously large) values of $Z_{\nk}$.  

In contrast to the case of purely electronic self--energies the QP approximation is known to lead to a too rough
description of the electron--phonon spectral function for low--energy electrons in the homogeneous electron gas\,(jellium). As
discussed by Engelsberg and Schrieffer\,\cite{engelsberg1963} this failure, 
although not as dramatic as the one found in the present case, 
is linked to the mixing of electronic and phononic excitations.  More precisely the authors identify three kind of excitations
that appear as poles of the Green's function when the energy of the electronic level is increased well above the Debye energy 
and the strength of the electron--phonon coupling is also increased. One is a purely QP state where the electron is dressed by 
a phonon cloud. The others lie in the continuum of electron--phonon pairs  composed by clothed electrons and clothed phonons being excited, 
with a constant momentum sum. 

%%%%%%%%%%%%%%%%%%%%%%%%%%%%%%%%%%%%%%%%%%%%%%%%%%%%%%%%%%%%%%%%%%%%%%%%%%%%%%%%%%%%%%%%%%%%%%%%%%%%%%%%%%%%%%%%%%
\section{Breakdown of the Quasi Particle approximation: the case of trans--polyacetylene and polyethylene}
\label{sec:brakdownQPapprox}
\begin{figure} [t]
\epsfig{figure=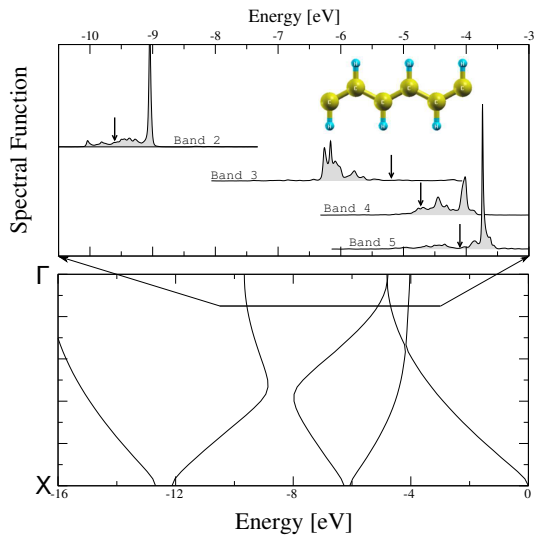,width=7.0cm}
\vspace{0.1cm}
\caption{\footnotesize{(color on line). The SFs of the last four occupied states in {\it trans}-polyacetylene (upper frame) are shown together
with the DFT--LDA bands (lower frame). The horizontal line in the bands frame represents the position of the {\kk}--point and the
energy range along which the SFs are displayed. The vertical arrows in the upper frame represents the energy position of the
unperturbed DFT levels.
Since these states correspond to in plane orbitals they are strongly affected
by the in plane atomic vibrations. The result is that the bare electronic levels are split in several polaronic
states.}}
\label{fig:Fig3}
\end{figure}

\begin{figure} [t]
\epsfig{figure=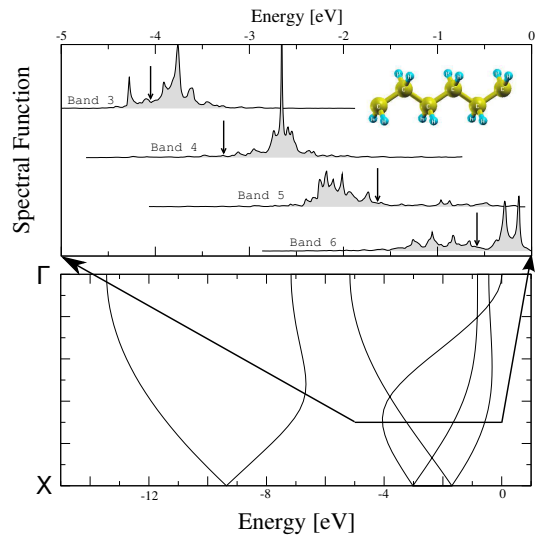,width=7.0cm}
\vspace{0.1cm}
\caption{\footnotesize{(color on line). Like in Fig.(\ref{fig:Fig3}) in the \tpe\,\,case.
}}
\label{fig:Fig4}
\end{figure}

%%%%%%%%%%%%%%%%%%%%%%%%%%%%%%%%%%%%%%%%%%%%%%%%%%%%%%%%%%%%%%%%%%%%%%%%%%%%%%%%%%%%%%%%%%%%%%%%%%%%%%%%%%%%%%%%%%
In the previous section we showed that the structures appearing in the SF of \tpa\,\,rule out any description of the 
coupled electron--phonon system in terms of QPs.
More importantly the rich structure of peaks appearing in the \Sf\,\,described in Fig.\,\ref{fig:SF_b1k1} is not a fortuitous case. 
It is actually a general trend both in \tpa\,\,and in \tpe. 
Indeed, in Figs.\,\ref{fig:Fig3} and \ref{fig:Fig4}, the bare electronic band structure and the corresponding \Sfs\,\, for a fixed $\kk$ vector
are shown in the upper frame. The position of the $\kk$ vector in the Brillouin Zone\,(BZ) is represented by an horizontal line in the lower frame
where the the valence bands of the two polymers are also reported. 

In the upper frame of  Figs.\,\ref{fig:Fig3} and \ref{fig:Fig4} there is also a sketch of the atomic structure of the 
polymers that helps to understand the main key differences among them. Both systems are linear polymers.
\Tpa\, is a conjugated polymer where each carbon atom forms four nearest--neighbour bonds. 
Three of the four carbon valence electrons are in $sp^2$ hybridized orbitals and two of the $\gs$-type bonds connect neighbour carbons along the 
one--dimensional backbone, while the third forms a bond with the hydrogen side group. 
\Tpe\,,  instead, is a $\gs$-bonded, non--conjugated polymer. The atoms in the unit cell do not lie on a single plane like in \tpa.
The C atoms are sp$^3$ hybridized and, as in the \tpe\, each C atom has four bonds. 

From the upper panels of  Figs.\,\ref{fig:Fig3} and \ref{fig:Fig4} it is evident that the EP interaction dramatically affects the
spectral functions. The most striking aspect is that the SFs exhibit a multiplicity of structures. Although in some cases a single and strong peak can
be observed the general trend is a very complex ensemble of peaks that makes impossible to apply the QP approximation.
We will give a more formal and mathematical description of the internal structure of these peaks in the next section. Here we would like to 
underline some of their general aspects.

A remarkable aspect is that some SFs are so largely structured that they span a large energy range, even $3\,eV$
(see the $6^{th}$ band of \tpe, Fig.\,\ref{fig:Fig4}). If they span a so large energy range, the SFs end up with 
overlapping each other in some cases (like the $4^{th}$ and $5^{th}$ band SFs in \tpa). 
The crucial and straightforward consequence is that it turns out difficult to associate a single and well defined 
energy to the electron and, more importantly, different bands will energetically merge pointing to a non trivial 
mixing of the electronic states.

When a SF covers a large energy range, one peak may be distant in energy from the others more than the Debye
energy ($0.4\,eV$ in \tpa\,\, and \tpe). For this reason each peak can not be simply interpreted 
as a main QP peak plus a phonon replica. This point will be further discussed in 
Appendix A  by using a
two band model.
Nevertheless  it is reasonable to speculate that
the formation of more than one peak suggests to reformulate the problem in a different framework where bare
electrons are mixed with phonons, and not simply screened by phonons. Each peak appearing in the SF would be
then identified by a mixed electron-phonon states. Since the \MB\, framework is not suitable to add 
information about the composition of the “new” mixed states, we will reformulate the problem in the next section 
by mapping the problem into and Hamiltonian representation. 


\begin{figure} [t]
\begin{center}
\epsfig{figure=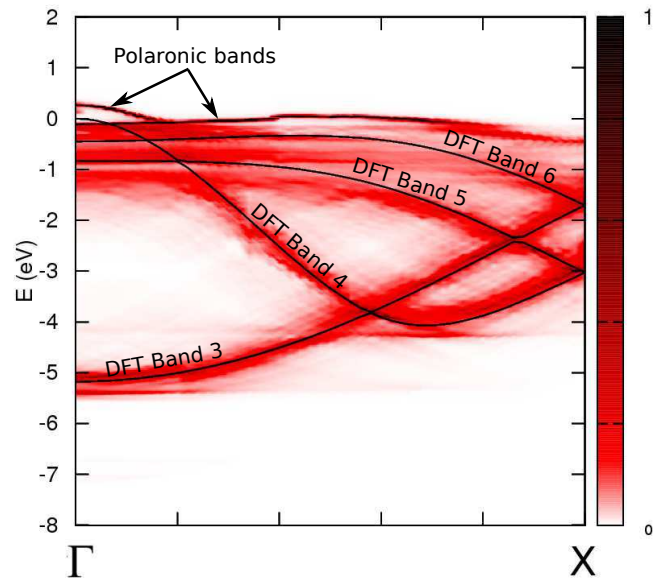,width=8.8cm} 
\end{center}
\vspace{-0.5cm}
\caption{\footnotesize{(color on line). Two dimensional plot of the spectral functions $\Delta Z_{\nk}(\go)$ for polyethylene
        in the last four occupied bands region. The DFT and the polaronic bands are opportunely labeled.
        In general the electronic levels acquire a large energy indetermination if compared to the DFT bands represented
        by solid black lines. The EP interaction moves up of about $300$\,meV the last two occupied bands leading to an increase of the band width.
        Moreover the 6$^{th}$ band near the $X$ point shows a large energy indetermination that makes it almost disappearing.}}
\label{fig:PA_Ank^2}
\end{figure}

A global view of the effect of the ZPM on the electronic structure of \tpe\,\,is given in Fig.\,\ref{fig:PA_Ank^2}
in the energy range of the last 4 occupied valence bands. The DFT electronic bands are drawn as a reference
of the electronic band structure before switching on the \epip

By defining $\Delta Z_{\nk}\(\go\)\equiv A_{\nk}\(\go\)\Delta \go$ the probability to find an electron $\mid \nk \ra$ in the small
energy range $\Delta \go = 50\,meV$, we made a bidimensional representation of the probability amplitude.
This is showed in Fig.\,\ref{fig:PA_Ank^2} by using a colored scale that goes from white (the less intense peak),
to black (the most intense one).
As a consequence this picture gathers all the information about the energy range covered by the \Sfs\,\,and the
intensity of all peaks.
In particular we observe that the $6^{th}$ band of \tpe\,\,moves up close to $\gC$-point and then the
electron completely disappears.

The resulting zero point renormalization of the gap at $\Gamma$ point of \tpe\,\,is $280 meV$, larger than the
trans-polyacetylene case~\cite{cannuccia}.
Such a difference is ascribed to the peculiar shape of the \tpa\, orbital at the $X$ point  whose $\pi$ character
corresponds to states perpendicular to the polymer axis. Thus they feel less the effect
of the in-plane vibrations. On the other hand for \tpe\,\,at $\Gamma$, the electrons are localized along the
$C-C$ bond, where the zero point motion effect of the electronic gap is sizable.
For what concerns the deeper states far from the gap, the effect of the electron-phonon coupling is equally strong.
In fact as they are in plane orbitals they are directly affected by in plane atomic vibrations.

We also observe that each band has a different energy width which evolves in different manners moving
from $\Gamma$ to $X$. An increasing of the bandwidth is normally associated to a consequent increase
of the delocalization of the orbitals. This fact can be used link the effect of the EP coupling 
to a increased electronic mobility mediated by the polaronic states.
%An enhancement of the mobility, opens new perspectives for future investigations and applications of polymers,
%in particular this picture can give a nice intriguing contribution to the debate on the origin of the mobility
%in polymers in terms of polaronic mobility.

In the next section we will go beyond this picture by introducing a general framework to link
the poles of the electron--phonon Green's functions to coupled packets of electron--phonon pairs.

%%%%%%%%%%%%%%%%%%%%%%%%%%%%%%%%%%%%%%%%%%%%%%%%%%%%%%%%%%%%%%%%%%%%%%%%%%%%%
\section{Internal structure of the polaronic states via an Hamiltonian representation}
\label{sec:frohlich}
%%%%%%%%%%%%%%%%%%%%%%%%%%%%%%%%%%%%%%%%%%%%%%%%%%%%%%%%%%%%%%%%%%%%%%%%%%%%%
In order to gain more insight into the complex structures that appear in the SFs of both \tpa\, and \tpe\, we propose, in this section,
a mapping of the Many--Body problem into an equivalent Hamiltonian representation. In this representation the
poles of the SF will appear as eigenvalues of a fictitious el--ph Hamiltonian. When this is solved in
a specific restricted sub--space of the entire Fock space, it will reproduce the same SF obtained by solving the 
Dyson equation within the Fan and Debye--Waller approximations for the self--energy.

In order to show this we start by rewriting the \Sf\,\,by using the well--known Lehmann representation\cite{mahan}:
\begin{align}
A_{\nk}\(\go,T\)=\sum_{I\kk} {\mid \la \Psi_0 \mid c^{\dg}_{\nk}\mid I\kk\(T\) \ra \mid}^2 
 \gd\(\go-E_{I\kk}\(T\)\),
\label{eq:H_1}
\end{align}
where $\mid I\kk\(T\) \ra$ are the true eigenstates of the system with energy $E_{I\kk}\(T\)$ which, in turns, represent the true
and real poles of the GF. As our initial Hamiltonian, Eq.\,(\ref{eq:sec_MBPT_1}), is composed of electrons and phonons the states
$\mid I\kk\(T\) \ra$ live an extended Fock space composed of electrons and phonons.

In the QP approximation the distribution of peaks appearing in Eq.\,(\ref{eq:H_1}) is approximated with a 
Lorentzian distribution centered at the QP energy with a width equal to the QP line-width. Thus,
Eq.\,(\ref{eq:H_1}) already underlines that 
the origin of the multiple poles in the \Sfs\,\,shown in Fig.\,\ref{fig:Fig3} and Fig.\,\ref{fig:Fig4}
is connected to the existence of more than one intense state $\mid I \kk \ra$ belonging to the same state $\mid \nk \ra$. 

Now we assume that Eq.\,(\ref{eq:H_1}) remains valid also when the exact self--energy is approximated by the Fan 
and Debye--Waller terms. And, in order to link the states $\mid I\kk\(T\) \ra$ to an Hamiltonian problem we start 
rewriting Eq.\,(\ref{eq:sec_MBPT_1}) in second quantization:
\begin{align}
\widehat H=\widehat H_{el}+\widehat H_{ph}+\widehat H_{el-ph},
\label{eq:H_2}
\end{align}
where $\widehat H_{el}$ is the electronic Hamiltonian, $\widehat H_{ph}$ is the independent phonons Hamiltonian 
and $\widehat H_{el-ph}$ is the \epi Hamiltonian. The last three terms, written in the second quantization, read
\begin{gather}
\widehat H_{el}=\sum_{\nk}\tilde{\gee}_{\nk}c^{\dagger}_{\nk}c_{\nk},
\label{eq:H_3}\\
\widehat H_{ph}=\sum_{\qq,\gl}\goql( b^{\dg}_{\qq \gl} \bql+\uot),
\tag{\ref{eq:H_3}$'$}\\
\widehat{H}_{el-ph}=\frac{1}{N_q}\sum_{\substack{n,n',\kk,\\\qq,\gl}}
             {g^{\qq,\gl}_{nn'\kk} c^{\dg}_{n\kk} c_{n'\kk-\qq}
             (b^{\dg}_{-\qq\gl}+b_{\qq\gl})}.
\tag{\ref{eq:H_3}$''$}
\end{gather}
In Eq.\,(\ref{eq:H_3}) $\tilde{\gee}_{\nk}$ is a single particle energy that we will shortly define.
$c^{\dg}_{n\kk}$ is the creation and $c_{n'\kk-\qq}$ is the annihilation electronic operators,
$b^{\dg}_{\qq \gl}$ and $\bql$ are the creation and
annihilation operators for phonons with energy $\goql$ and wave vector $\qq$. $g^{\qq,\gl}_{nn'\kk}$ 
are the EP coupling matrix elements (see Eq.\,(\ref{eq:sec_diagrams_4})). 

We want now to use Eqs.\,(\ref{eq:H_3}) to calculate the states $|I\kk\(T\)\ra$.
To this end we note, from the (a) frame of Fig.\ref{fig:sec_diagrams_1}, that the Fan \se 
makes an initial state $|n\kk\ra$ to scatter with a phonon state $|\qq\gl\ra$, with population $N_{\qq\gl}\pm 1$, in a final state 
$|n'\kk-\qq\ra$. Only one phonon is exchanged and in the self--energy loop the intermediate states 
are the composite pairs $ \mid n' \kk-\qq \ra \otimes \mid N_{\qq\gl}\pm 1 \ra$ with energy 
$\gee_{n'\kk-\qq}\pm \go_{\qq\gl}$. This energy, indeed, appears in the denominator of Eq.\,(\ref{eq:sec_diagrams_10}).

Physically this means that, if we introduce the general state product of an electronic and phononic part
$\mid \npkmq \ra \otimes \mid N_{\qq\gl} \pm 1 \ra$, the intermediate states of the self--energy are all possible 
combinations with different $\qq$ and $\gl$. It follows that we can guess, at a given temperature,
\begin{multline}
\mid I{\kk}\(T\) \ra= \sum_n A^I_{\nk}\(T\) \mid \nk \ra + \\ + \sum_{n'\qq\gl} B^{I\gl}_{n'\kk-\qq}\(T\)
                 \mid \npkmq \ra \otimes \mid N_{\qq\gl}\(T\)\pm 1 \ra.
\label{eq:H_4}
\end{multline}
The coefficients $A^I$ and $B^{I\gl}$ can be found by diagonalizing Eq.\,(\ref{eq:H_2}) in the space of electron--phonon states spanned
by the definition given in Eq.\,(\ref{eq:H_4}). More precisely,
in order to expand the matrix form of Eq.\,(\ref{eq:H_2}) we notice that, due to Eq.\,(\ref{eq:H_4})
the basis set will be composed of the following elements
\begin{align}
\mid \nk \ra \mid N_{\qq\gl} \ra,\,\,\mid n\kk-\qq \ra \mid N_{\qq\gl}\pm 1 \ra.
\end{align}
At zero temperature the basis set is reduced to
\begin{align}
\mid \nk \ra \mid 0_{ph}\ra,\,\,\mid n\kk-\qq \ra \mid 1_{\qq\gl} \ra,
\end{align}
and it reflects the fact that at $T=0\,K$ there are no phonons in the ground state. In this restricted basis
the Hamiltonian reads
\begin{align}
\widehat H =
\begin{bmatrix} 
\ddots &                        &        &                            &        \\ 
       & \tilde{\gee}_{\nk}\delta_{nn'} &        &  H^{ep}_{nn'\qq\gl}        &        \\ 
       &                        & \ddots &                            &        \\
       & (H^{ep}_{nn'\qq\gl})^{\dagger}  &        & \tilde{\gee}_{\nk-\qq}\gd_{nn'}+\goql \gd_{\qq\qq'}\gd_{\gl\gl'} & \\
       &                        &        &                            & \ddots \\
\end{bmatrix}
.
\label{eq:H_5}
\end{align}
The fact that $\hat{H}_{el-ph}$ is Hermitian makes also $\hat{H}$ Hermitian. The equivalence of Eq.\,(\ref{eq:H_1}) with the spectral function calculated in
Sec.\ref{sec:DynamicalSEeffects} 
is obtained by imposing  $\tilde{\gee}_{n\kk}=\gee_{n\kk}+\Sigma^{DW}_{n\kk}$. This equivalence is 
proved analytically, in the zero temperature limit, in Appendix~\ref{appendixA} for a simple two--levels model.

From Eq.\,(\ref{eq:H_5}) we also note that the number of states $\mid I{\kk}\(T\)\ra$ is equal to the
dimension of the matrix and it is obtained by multiplying the number of electronic bands times the number 
of $\qq$ vectors times the number of phononic branches, $\gl$. 
As a consequence the number of states $\mid I{\kk}\(T\)\ra$ is larger than that of $\widehat H_{el}$.  This confirms the fact that, in
Eq.\,(\ref{eq:H_1}), the states $|I\kk\(T\)\ra$ form a continuum that, in the QP approximation, dresses
the bare electronic state. This dressing is, {\it de--facto}, represented by a cloud of mixed electron--phonon states
surrounding the QP energy with a Lorentzian distribution.

\begin{figure} [t]
\begin{center}
\vspace{-0.3cm}
\epsfig{figure=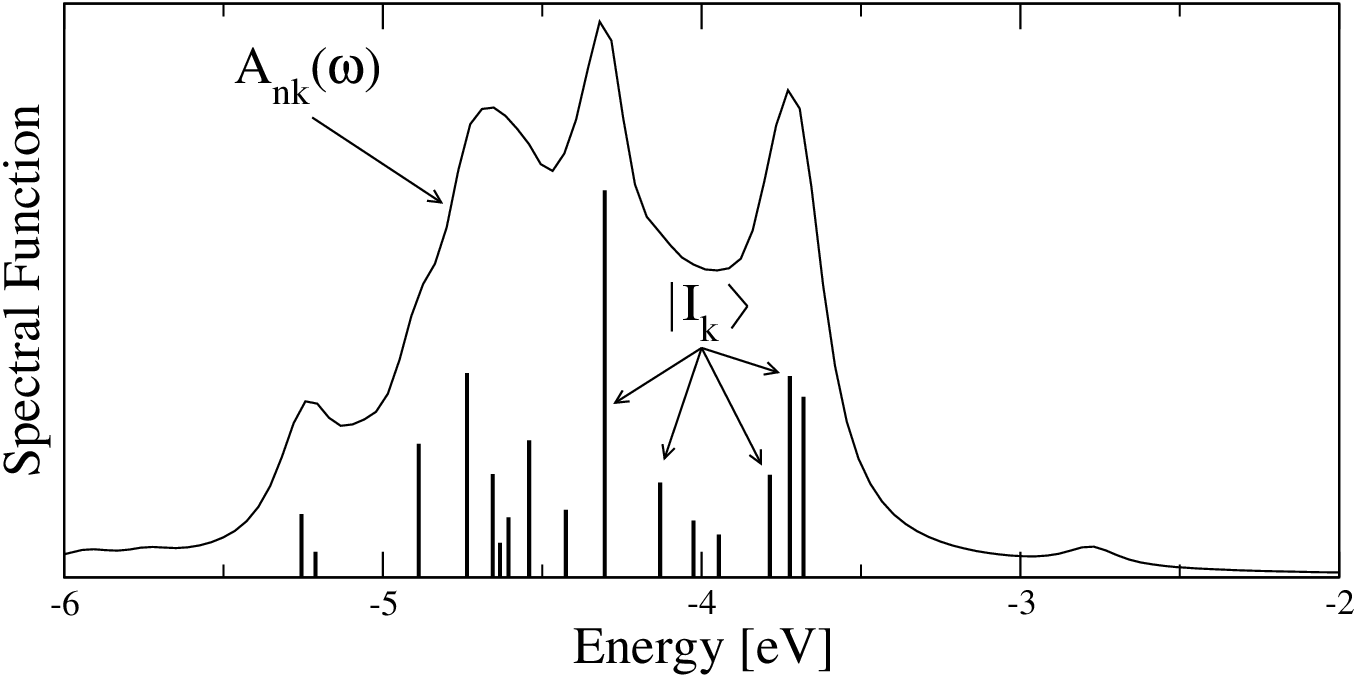,width=7.0cm,clip=,bbllx=61,bblly=197,bburx=712,bbury=522} 
\end{center}
\vspace{-0.7cm}
\caption{\footnotesize{\Tpa. The \Sf\,\,of the state $\mid n=4, \kk=0.2(\frac{2\pi}{a},0,0)\ra$ is decomposed in polaronic states, each 
labeled by $\mid I \kk \ra$. Several structures appear thus ruling out the QP approximation.}}
\label{fig:band4_k1_Impulsi}
\end{figure}

The Hamiltonian $\widehat H$ (Eq.\,(\ref{eq:H_5})) is diagonalized for \tpa\,\,including 
$30$ electronic bands, $10\,\qq$-vectors and $12$ phonon branches.
Since the Hamiltonian is written on a complete basis, the coefficients $A^I_{\nk}$ and $B^{I\gl}_{n'\kk-\qq}$ 
satisfy the following condition
\begin{align}
\sum_{n} \mid A^{I}_{\nk}\mid^2 + \sum_{n'\qq\gl} \mid B^{I\gl}_{n'\kk-\qq} \mid^2 =1,
\label{eq:H_6}
\end{align}
which ensures that the spectral function $A_{n\kk}\(\go,T\)$ is correctly normalized to 1 when integrated over all
frequencies.
Once the eigenstates $\mid I{\kk} \ra$ and the eigenvalues $E_{I{\kk}}$ are known,
the \Sf\,\,can be calculated according to Eq.\,(\ref{eq:H_1}) and all
the peaks appearing in the \Sfs\,\,of state $\mid \nk \ra$, are unambiguously
labeled with a particular 
$\mid I{\kk} \ra$ state, having $\mid \nk \ra$ as the pure electronic component.
Let us consider the $\mid n=4, \kk=0.2(\frac{2\pi}{a},0,0)\ra$ state of \tpa\,\,as an
example. In Fig.\,\ref{fig:band4_k1_Impulsi} it is shown the corresponding zero temperature \Sf.
The poles and the corresponding residuals are indicated in Fig.\,\ref{fig:band4_k1_Impulsi} by bars
with different heights.
The residuals are given by $\mid A^{I}_{\nk} \mid^2$, that is the probability to find the polaronic state in the pure electronic 
$\mid \nk \ra$ state. This reminds the physical meaning of the $Z_{\nk}$ factors, and we 
use this similarity to define $Z^{I}_{\nk}=\mid A^{I}_{\nk} \mid^2$.
Nevertheless from Eq.\,(\ref{eq:H_4}) it is evident that
the smaller the $Z^{I}_{\nk}$ is, the less the polaronic state can be assimilated to
an electron. It means that the mixed \ep\,contribution in Eq.\,(\ref{eq:H_4}) indeed weights the most.
In the case of Fig.\,\ref{fig:band4_k1_Impulsi} the $\mid A^{I}_{\nk} \mid^2$ can even be as small as 0.2.

\begin{figure} [t]
\begin{center}
\vspace{0.2cm}
\epsfig{figure=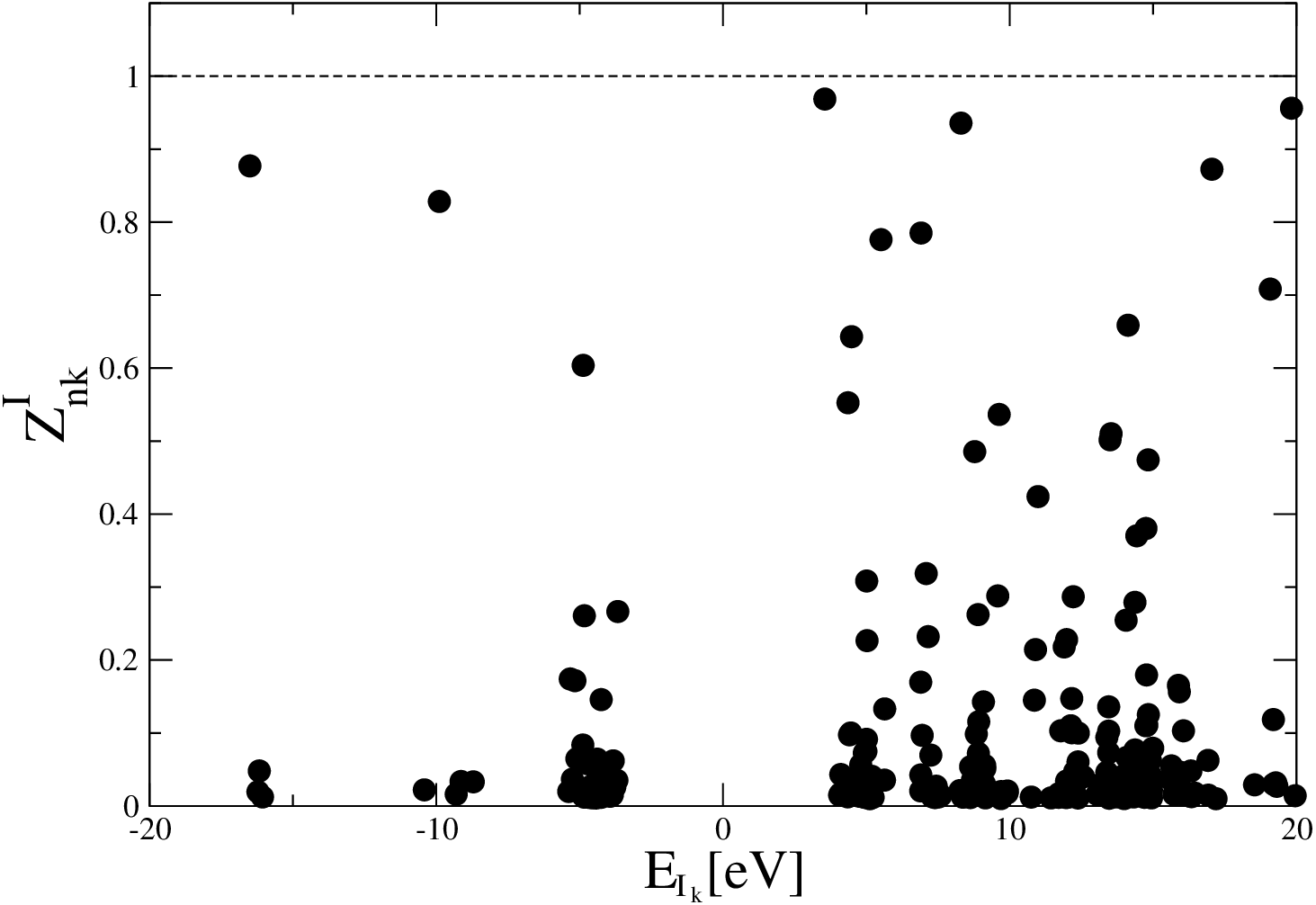,width=7.0cm,clip=,bbllx=8,bblly=39,bburx=714,bbury=522} 
\end{center}
\vspace{-0.7cm}
\caption{\footnotesize{\Tpa. The projection of the
polaronic states over the corresponding pure electronic state, $Z^I_{\nk}$ is shown. The dashed line represents 
$Z^I_{\nk}$ of a pure electron state as a reference value.}}
\label{fig:Charge_vs_Epolaronic}
\end{figure}

These small values of $Z^{I}_{\nk}$  represent a general trend. In Fig.\,\ref{fig:Charge_vs_Epolaronic} $Z^{I}_{\nk}$
is plotted as a function of the polaronic eigenvalues.
It can be noted that only few polaronic states have $Z^I_{\nk}\simeq1$. Most of all are below $0.5$, instead.
It means that the mixed EP part of the eigenstate, shown in Eq.\,(\ref{eq:H_4}),
plays a dominant role.

A small value of $Z^I_{\nk}$ points to a non trivial physical property of the polaronic states. When 
$Z^I_{\nk}\rightarrow 0$ it is evident that only the mixed terms in the sum, where the electrons and the phonons appear together, 
are non zero. This means that the electrons cannot move in the system alone, even if dressed by an electron--phonon cloud, but
need to build up true bound electron--phonon states. This is a clear fingerprint of the breakdown of the QP approximation.

The generic definition of the polaronic state, Eq.\,(\ref{eq:H_4}), allows to calculate the mean value of any observable that lives
in the mixed electron--phonon space. 
For example we can evaluate the matrix elements of the atomic indetermination operator 
as follows
\begin{multline}
u^2_{\ga I\kk s }\equiv\la I\kk \mid u^2_{\ga,s} \mid I\kk \ra = \\ 
\sum_{\qq\gl}\(\frac{1}{2N_q\goql M_s}\) \gee_{\ga}\(\qq\gl/s\)\gee^{\ast}_{\ga}\(\qq\gl/s\)\times\\
\times\[\sum_{n} \mid A^{I}_{\nk}\mid^2 + 3  \sum_{n'}\mid B^{I\gl}_{n'\kk-\qq} \mid^2 \].
\label{eq:H_8}
\end{multline}
By using these $u^2_{\ga I \kk s }$ we can associate an average quantum size to the atoms.
\begin{table}[t]
\begin{center} 
\begin{tabular}{|c|c|c|c|c|}   \hline 
\multicolumn{1}{|r|}{}&\multicolumn{2}{|r|}{\Tpa}&\multicolumn{2}{c|}{\tpe} \\ \hline
         &   $  C  $  &  $  H  $ &   $  C  $  &  $  H  $   \\ 
         &   $a.u. $  &  $ a.u.$ &   $a.u. $  &  $ a.u.$   \\ \hline 
$\hat x$ &   $0.18 $  &  $0.55 $ &   $0.1  $  &  $0.32 $   \\ \hline
$\hat y$ &   $0.13 $  &  $0.36 $ &   $0.07 $  &  $0.21 $   \\ \hline
$\hat z$ &   $0.11 $  &  $0.56 $ &   $0.07 $  &  $0.34 $   \\ \hline  
\end{tabular}
\caption{\footnotesize{Atomic amplitudes obtained by evaluating the matrix elements of operator
$\sqrt {\mathbf u^2_{I\kk s}}$. Because of its smaller mass the hydrogen quantum atomic size is three times larger than the carbon atom one.
Nevertheless the constraint imposed by the different geometries makes the deviation of the two 
polymers appreciably different.}}
\label{tab:AtomIndetermination}
\end{center}
\end{table}
The values for the $C$ and $H$ species, calculated by Eq.\,(\ref{eq:H_8}), are shown 
in Tab.\,\ref{tab:AtomIndetermination}. These values point to the fact that 
the atoms acquire an indetermination larger along the polymer axis.
Since $H$ is lighter than $C$ the atomic quantum size is larger. The different constraint created
by the geometries is the cause of the different $\sqrt {u^2_{\ga I \kk s}}$ values between the two polymers.

The values of the atomic indetermination suggest
that electrons and phonons exert a cooperative effect on each other. The
charge density spreads all along the polymer, while the atoms squeeze along $\hat y$
and $\hat z$ directions, widening along $\hat x$.
This cooperation can cause, for example, an enhancement of the mobility, opening therefore new
perspectives for future investigations and applications of polymers.

%%%%%%%%%%%%%%%%%%%%%%%%%%%%%%%%%%%%%%%%%%%%%%%%%%%%%%%%%%%%%%%%%%%%%%%%%%%%%%%%%%%%%%%%%%%%%%%%%%%%%%%%%%%%%
\section{Conclusions}
\label{sec:conclusions}
%%%%%%%%%%%%%%%%%%%%%%%%%%%%%%%%%%%%%%%%%%%%%%%%%%%%%%%%%%%%%%%%%%%%%%%%%%%%%%%%%%%%%%%%%%%%%%%%%%%%%%%%%%%%%
In this work we have shown that the Heine--Allen--Cardona approach suffers of some severe limitations when applied to
predict the zero temperature energy correction in low dimensional systems. 
We extensively describe a fully dynamical extension of the Heine--Allen--Cardona approach
to show that the  zero point motion effect severely questions the reliability of the QP picture in {\it trans}-polyacetylene
and polyethylene.

The single particle spectral functions, indeed, exhibit multiple structures at $T=0\,K$.
The formation of additional structures caused by the strong electron--phonon interaction
is interpreted in terms of composed electron--phonon states, what we call in this work polaronic states.
These states are precisely defined by
mapping the structures of the Many--Body spectral functions into the solution of an eigenvalue problem.

Thanks to this important mapping the non 
perturbative nature of the polaronic states appears as a coherent superposition 
of electron--phonon pairs. And the cooperative dynamics between electrons and atoms in these states rules out
any description in terms of bare atoms and quasiparticles. 

The resulting coupled electronic and atomic dynamics pave the way for new investigations
in polymers and more in general in low dimensional nanostructures.
The cooperative dynamics of electrons and phonons in the polaronic states can have potential
physical implications, as for example, an enhancement of the electronic mobility.

More generally the breakdown of the quasiparticle picture imposes a critical
analysis of the previous results obtained using purely electronic theories.

\appendix
%%%%%%%%%%%%%%%%%%%%%%%%%%%%%%%%%%%%%%%%%%%%%%%%%%%%%%%%%%%%%%%%%%%%%%%%%%%%%%%%%%%%%%%%%%%%%%%%%%%%%%%%%%%%%
\section{A two--levels model to verify the Hamiltonian representation}
\label{appendixA}
In Sec.\ref{sec:frohlich} we have given physical arguments to support the choice of a specific
limited Fock space where the Hamiltonian problem is solved. 
These arguments were guided by the final goal of introducing an Hamiltonian representation that 
gives exactly the Green's functions corresponding to the Fan approximation for the self--energy.
To better investigate and confirm this ansatz
let us consider two levels of energies $\gee_i=0,E$ coupled to a phonon of energy 
$\go_0$ at $T=0\,K$.

Eqs.\,(\ref{eq:H_2}-\ref{eq:H_5}) thus reduce to a simple expression for the Hamiltonian
of this system
\begin{align}
H=\sum_{i=1}^{2} \epsilon_{i}c^{\dagger}_{i}c_{i}+\omega_{0}b^{\dagger}_{0}b_{0}+
\sum_{\substack{i=1,2 \\ j=1,2 \\ i \neq j}} g c^{\dagger}_{j}c_{i}(b^{\dagger}_{0}+b_{0}).
\end{align}
As discussed in Sec.\ref{sec:frohlich} we consider for the finite basis the following ansatz:
\begin{align}
\mid 1 \rangle \mid 0_{ph} \rangle,\mid 2 \rangle \mid 0_{ph} \rangle,\\
\mid 1 \rangle \mid 1_{ph} \rangle,\mid 2 \rangle \mid 1_{ph} \rangle,
\end{align}
where by $\mid i \ra$ we mean an electron in the level $i^{th}$.
The dimension of the Hamiltonian matrix is then given by multiplying 2 bands $\times\,1\,\qq$ point
$\times\,1$ phonon branch, resulting in a $4\times4$ matrix
\begin{align}
  {\mathcal H}=\left ( 
  \begin{array}{cccc}
  0 & 0 & 0          & g \\ 
  0 & E & g          & 0 \\ 
  0 & g & \go_{0} & 0 \\
  g & 0 & 0          & E+\go_{0}
  \end{array}
  \right),
\end{align}
which can be diagonalized in two blocks, obtaining the following four energy levels
\begin{align}
E_{1}=\frac{E+\go_0 + \sqrt{(E+\go_0)^2 + 4g^2}}{2},
\label{eq:eigen1}
\end{align}
\begin{align}
E_{2}=\frac{E+\go_0 - \sqrt{(E+\go_0)^2 + 4g^2}}{2},
\label{eq:eigen2}
\end{align}
\begin{align}
E_{3}=\frac{E+\go_0 + \sqrt{(E-\go_0)^2 + 4g^2}}{2},
\label{eq:eigen3}
\end{align}
\begin{align}
E_{4}=\frac{E+\go_0 - \sqrt{(E-\go_0)^2 + 4g^2}}{2}.
\label{eq:eigen4}
\end{align}
The corresponding four eigenvectors are
\begin{align} 
\mid I_{1} \ra=\frac{1}{N_1} \( \frac{g}{E_1},\,0\,,0,\,1 \), \\
\mid I_{2} \ra=\frac{1}{N_2} \( 1,\,0\,,0,-\frac{g}{E_1}  \), \\
\mid I_{3} \ra=\frac{1}{N_3} \( 0,\,1,\,-\frac{g}{\go_0-E_3},\,0 \), \\
\mid I_{4} \ra=\frac{1}{N_4} \( 0,\frac{g}{\go_0-E_3},\,1\,,0 \), 
\end{align}
where $N_i$ are the normalization factors, with $N_1=N_2$ and $N_3=N_4$.

These are the needed ingredients to calculate the \GFs\,as matrix element of the resolvent
\begin{align}
{\cal{G}}_i(\go)=\la vac \mid c_1\, \frac{1}{\go-H}\, c^{+}_1 \mid vac \ra,
\label{eq:G_resolvent1}
\end{align}
where $\mid vac \ra$ is the vacuum of phonons and electrons.
Expanding in eigenstates of the system, Eq.\,(\ref{eq:G_resolvent1}) becomes
\begin{eqnarray}
{\cal{G}}_i(\go)&=&\sum_{j=1}^{4}\la vac \mid c_i\, \frac{1}{\go-H}\mid I_j\ra\la I_j\mid c^{+}_i \mid vac \ra\nonumber \\
&=&\sum_{j=1}^{4}{\mid \la vac \mid c_i \mid I_j \ra \mid}^2 \frac{1}{\go-E_{j}}.
\label{eq:G_resolvent2}
\end{eqnarray}
In order to show that the \Sfs\,\,calculated from Eq.\,(\ref{eq:G_resolvent2}) are equivalent to the ones obtained
in the MB approach, the \GF\,\,for the $1^{st}$ state is evaluated from Eq.\,(\ref{eq:G_resolvent2}) as follows
\begin{align}
{\cal{G}}_1(\go)=\frac{1}{E_1-E_2} \[\frac{-E_2}{\go-E_1}+\frac{E_1}{\go-E_2}\].
\label{eq:Ham_GF_2bands}
\end{align}
On the other hand the Fan approximation for the \sev Eq.\,(\ref{eq:sec_diagrams_10}), in this test case reduces to
\begin{multline} 
\gS^{Fan}_1(\go)=g^2  \[\frac{N(\go_0)+1-f_2}{\go-E-\go_0-i0^+}+\capo \frac{N(\go_0)+f_2}{\go-E+\go_0-i0^+}\]. 
\label{eq:Fan_se_2bands}
\end{multline} 
At zero temperature the Bose occupation factors vanish. The Fermi occupation factor $f_2$ is zero
because the level is empty and
Eq.\,(\ref{eq:Fan_se_2bands}) becomes 
\begin{align} 
{\cal{G}}_1(\go)=\frac{1}{\go-g^2 \[\frac{1}{\go-E-\go_0}\]-i0^+}. 
\label{eq:Fan_GF_2bands}
\end{align}
The poles of Eq.\,(\ref{eq:Fan_GF_2bands}) are $\go=E_1$ and $\go=E_2$, defined by Eqs.\,(\ref{eq:eigen1}--\ref{eq:eigen2}). 
The residues evaluated at each pole are shown as follows
\begin{eqnarray}
    \go= E_1 \,\,\,\,\,\,  &\text{Res}_1&=-\frac{E_2}{E_1-E_2},
\label{eq:res_E1}\\
    \go= E_2 \,\,\,\,\,\,  &\text{Res}_2&= \frac{E_1}{E_1-E_2}.
\label{eq:res_E2}
\end{eqnarray}
The final expression for ${\cal{G}}_1$ in the \MB approach is then
\begin{align} 
{\cal{G}}_1(\go)=\frac{1}{E_1-E_2} \[-\frac{E_2}{\go-E_1-i0^+}+\frac{E_1}{\go-E_2-i0^+}\], 
\label{eq:Fan_GF_2bands_final}
\end{align}
that is equivalent to Eq.\,(\ref{eq:Ham_GF_2bands}). 

From Eqs.\ref{eq:eigen1}--\ref{eq:eigen2} we notice that 
\begin{align} 
E_1-E_2=\sqrt{(E+\go_0)^2+4g^2},
\label{eq:DeltaEnergy}
\end{align} 
which is larger than $\go_0$ and it is also larger then the same energy difference when $g\rar0$.
This clearly means that both electrons take part in the formation of the polaronic state thanks to the additional energy provided by the 
\ep\, coupling.
This also implies that each additional structure cannot be interpreted in energetic terms as simply
an electron ``plus'' one phonon.

%%%%%%%%%%%%%%%%%%%%%%%%%%%%%%%%%%%%%%%%%%%%%%%%%%%%%%%%%%%%%%%%%%%%%%%%%%%%%%%%%%%%%%%%%%%%%%%%%%%%%%%%%%%%%
\section{Calculation Details}
\label{appendixB}
The phonon modes and the electron--phonon matrix elements were calculated using a
uniform grid of $10 \times 1 \times 1$ k--points.
We used a plane--waves basis and norm conserving pseudo-potentials\,\cite{tro-martin} for the 
carbon and hydrogen atoms. The exchange correlation potential has been treated within the local density approximation.
For the ground--state calculations we used the \texttt{PWSCF} code\,\cite{Giannozzi2009}.
The Fan self-energy and the Debye--Waller contribution are calculated using a random grid of transferred momenta,
using the \texttt{yambo} code\,\cite{Marini20091392}.

The numerical evaluation of Eq.\,(\ref{eq:sec_diagrams_10}) is a formidable task. Indeed the use of a fine sampling 
of the BZ is prohibitive. The reason is that such large grids of transferred momenta are inevitably connected
with the use of equally large grids of $\kk$--points. An alternative solution, that we used in the present
calculations, is to fix a certain  $\kk$--points grid and to perform the integration of the BZ by using a random 
grid of points to perform the $\qq$ summation in Eq.\,(\ref{eq:sec_diagrams_10}).

Moreover, in order to speed--up the convergence with the number of random points and to take in account the divergence at 
$\qq\rightarrow\,0$ of the  $|g^{\gql}_{n' n \kk}|^2$  matrix elements we divide the BZ in small spherical regions
$R_{\qq}$ centered around  each $\qq$ point.
Eq.\,(\ref{eq:sec_diagrams_10}) can be then rewritten as
\begin{multline}
    \Sigma^{Fan}_{n\kk}\(\go,T\) =  
\sum_{\QQ} 
\sum_{n'\gl} 
\(\int_{R_{\QQ}}\,d\qq \frac {\gsq}{\Omega_{RL}}\)  \\
\[ \frac{N_{\QQ\gl}\(T\)+1-f_{n'\kk-\QQ}}{\go-\gee_{n' \kk-\QQ} -\goql -i0^{+}} \right. \\
+ \left. \frac{N_{\QQ\gl}\(T\)+f_{n'\kk-\QQ}}{\go-\gee_{n' \kk-\QQ}+\goql -i0^{+}}\].
\label{eq:App_2}
\end{multline}
In Eq.\ref{eq:App_2} the $d\qq$ integral is calculated using a numerical Montecarlo technique and taking explicitly into account
the $\qq\rightarrow\,0$ divergence of $|g^{\gql}_{nn' \kk}|^2$:
\begin{align}
\int_{R_{\QQ}}\,d\qq \frac {\gsq}{\Omega_{RL}}\approx \frac {\mid \QQ \mid^2 {\mid  g^{\QQ\gl}_{n n' \kk} \mid}^2}{\Omega_RL} 
\(\int_{R_{\QQ}}\,d\qq \qq^{-2}\).
\label{eq:App_3}
\end{align}
In Eq.\ref{eq:App_3} the three--dimensional $\qq$ integration compensates the $\qq^{-2}$ divergence making the 
numerical evaluation of Eq.\ref{eq:App_2} feasible. Moreover, while  ${\mid  g^{\QQ\gl}_{n n' \kk} \mid}^2$ diverges as
$\mid\QQ\mid^{-1}$, $\mid \QQ \mid^2 {\mid  g^{\QQ\gl}_{n n' \kk} \mid}^2$ is regular when $\QQ\rightarrow 0$.

\section*{Acknowledgments}
%%%%%%%%%%%%%%%%%%%%%%%%%%%%%%%%%%%%%%%%%%%%%%%%%%%%%%%%%%%%%%%%%%%%%%%%%%%%%%%%%%%%%%%%%%%%%%%%%%%%%%%%%%%%%
Financial support was provided by the European Research Council Advanced Grant DYNamo (ERC-2010-AdG -Proposal No. 267374), 
Spanish (FIS2011-65702-C02-01 and PIB2010US-00652 ), ACI-Promociona (ACI2009-1036), Grupos Consolidados UPV/EHU del 
Gobierno Vasco (IT-319-07) and  by the {\em Futuro in Ricerca} grant No. RBFR12SW0J of the
Italian Ministry of Education, University and Research.
Computational time was granted by i2basque, SGIker Arina,  BSC “Red Espanola de Supercomputacion” 
and the CASPUR computational resources (Italy).

\bibliographystyle{unsrtnat}
\bibliography{prb}	

\end{document}

%% file: alias.tex
\newcommand{\be}{\begin{equation}}
\newcommand{\ee}{\end{equation}}
\newcommand{\bea}{\begin{eqnarray}}
\newcommand{\eea}{\end{eqnarray}}

\def\pp		{{\bf p}}
\def\rr		{{\bf r}}

\def\QQ		{{\bf Q}}

\def\RR		{{\bf R}}

\def\qq		{{\bf q}}
\def\kk		{{\bf k}}

\def\ga         {\alpha}
\def\gb         {\beta}

\def\gC         {\Gamma}
\def\gd         {\delta}

\def\gee        {\epsilon}
\def\gl         {\lambda}

\def\go         {\omega}

\def\goql       {\omega_{\qq \gl}}
\def\gql        {\qq \gl}
\def\gon        {\omega_{n}}
\def\goj        {\omega_{j}}
\def\goi        {\omega_{i}}

\def\igoi       {i\omega_{i}}
\def\igoj       {i\omega_{j}}
\def\gr         {\rho}
\def\gs         {\sigma}
\def\gS         {\Sigma}

\def\capo       {\right.\\ \left.}

\def\rar        {\rightarrow}  
\def\la         {\langle}
\def\ra         {\rangle}
\def\dg         {\dagger}

\def\uob        {\frac {1}{\gb}}

\def\uot        {\frac {1}{2}}

\def\uon        {\frac {1}{N_{q}}}

\renewcommand{\[}{\left[}
\renewcommand{\]}{\right]}
\renewcommand{\(}{\left(}
\renewcommand{\)}{\right)}
\def\MBPT       {many-body perturbation theory}
\def\MB         {many-body }

\def\GF         {Green's function }

\def\GFs         {Green's functions }

\def\tpa        {{\it trans}--polyacetylene}
\def\Tpa        {{\it Trans}--polyacetylene}
\def\tpe        {polyethylene}
\def\Tpe        {Polyethylene}

\def\Vscf       {\widehat{V}_{scf} }

\def\epc        {electron--phonon coupling }
\def\epcp        {electron--phonon coupling. }

\def\epi        {electron--phonon interaction }
\def\epip        {electron--phonon interaction.}

\def\ep         {electron--phonon }
\def\dw         {Debye--Waller }
\def\se         {self--energy }

\def\sep         {self--energy. }
\def\sev         {self--energy, }

\def\nk         {n{\bf k}}

\def\npkmq      {n' {\bf k-q}}

\def\bql         {\hat{b}_{\qq \gl}}

\def\bdql        {\hat{b}^{\dg}_{-\qq \gl}}

\def\gsq         {{\mid g^{\gql}_{n n' \kk} \mid}^2}
\def\Sf          {spectral function}
\def\Sfs         {spectral functions}

\def\bverb      {\renewcommand{\baselinestretch}{1}\begin{verbatim}}
\def\everb  	{\end{verbatim}\renewcommand{\baselinestretch}{1.3}}